\documentclass[12pt, a4paper,notitlepage]{article}
\input epsf

\scrollmode
\def\be{\begin{equation}}
\def\ee{\end{equation}}

\begin{document}

\title{\large A modified one-dimensional Sznajd model}
\author{\normalsize Juan R. S\'anchez \\
\\
\normalsize Facultad de Ingenier\'\i{}a \\
\normalsize Universidad Nacional de Mar del Plata \\
\normalsize J.B. Justo 4302, Mar del Plata, Argentina. \\
\normalsize \it jsanchez@fi.mdp.edu.ar}
\date{}

\maketitle


{\small {\bf Abstract.}
The Sznajd model is an Ising spin model representing a simple mechanism of making up decisions in a closed community. In the model each member of the community can take two attitudes A or B represented by a spin up or spin down state respectively. 
It has been shown that, in one-dimension
starting from a totally random initial state, three final fixed points can be
obtained; all spins up, all spins down or an {\it antiferromagnetic} state in which each site
take a state which is opposite from its two nearest neighbors. Here, a modification of the
updating rule of the Sznajd model is proposed in order to avoid such antiferromagnetic state since it is considered to be an {\it unrealistic} state in a real community. }

\vspace{0.5cm}
The Sznajd model is a successful Ising spin model describing a simple mechanism of making up decisions in a closed community. The model allow each member of the community to have two attitudes, to vote for option A or to vote for option B. These two attitudes are identified with the state of spins variables up or down respectively. A dynamic is established in the model in which a selected pair of adjacent spins influence their nearest neighbors through certain rules, applied in a random sequential manner.  
In several votes (units of evolution time) some difference $m$ of voters for A and against is expected.
The dynamic rules of the Sznajd model are~\cite{1}

\vspace{0.2cm}
-- if $S_iS_{i+1}=1$ then $S_{i-1}$ and $S_{i+2}$ take the
direction of the selected pair [i,i+1],  \hspace\fill  ($r1$)

-- if $S_iS_{i+1}=$ -$1$ then $S_{i-1}$ takes the direction of
$S_{i+1}$ and $S_{i+2}$ the direction of $S_i$, \hspace\fill($r2$)

\vspace{0.2cm}
\noindent being $S_i$ the state of the spin variable at site $i$.
These rules describe the influence of a given pair of members of the community on the decision of its nearest neighbours.

In one dimension, the original rules give rise to three limiting cases in the
evolution of the system

\vspace{0.2cm}
(i)
all members of the community vote for $A$ (all spins up),

(ii)
all members of the community vote for $B$ (all spins down),

(iii) $50$\% vote for $A$ and $50$\% vote for $B$ ({\it alternating} state).

\vspace{0.2cm}
Here, attention is paid to the last limiting antiferromagnetic case (iii). This antiferromagnetic case, although posible in other spins systems, can be considered to be quite {\it unrealistic} in a model trying to represent the behavior of a community.
To achieve {\bf exactly} a $50$-$50$ final state in a community is almost impossible, specially if it is composed by more than a few dozens of members.~\cite{2}
On the other hand such antiferromagnetic state implies that each member of the 
community is surrounded by a neighbor which has an opposite opinion. 
A quite ``uncomfortable" situation, certainly.

From a simulational point of view, if the evolution of a one-dimensional Sznajd model 
is {\bf started} from an antiferromagnetic state, i.e., a chain of neighbors with opposite opinions, the original dynamic rules does not give rise to any evolution at all. 

\vspace{0.2cm}
In order to avoid the unrealistic $50$-$50$ alternating final state, new dynamic rules are proposed:

\vspace{0.2cm}
-- if $S_i S_{i+1} = 1$ then $S_{i-1}$ and  $S_{i+2}$ take the same direction of
	the pair $[i,i+1]$, \hspace\fill ($r1$) 
	
-- if $S_i S_{i+1} =$ -$1$ then $S_{i}$ take the direction of $S_{i-1}$ and $S_{i+1}$
	take the direction of $S_{i+2}$. \hspace\fill ($r2$)

\vspace{0.2cm}
Using the new rules, in case of disagreement of the
pair $S_i$-$S_{i+1}$, rule $r2$ make the spin $i$ to ``feel more
confortable" since it ends up with at least one neighbor having its own opinion.  

Two samples of evolution of a system following the new rules and starting from 
an antiferromagnetic state are shown in Fig. 1, for a $N=100$ lattice size. 
It can be seen that the $50$-$50$ final
state in completely avoided and that the other two types of total agreement 
(ferromagnetic) final states
can be achieved, with equal probability, starting the systems from the same initial condition. 
Time $t$ is advanced by one when each spin of the lattice has had one
(probabilistic) opportunity to be updated. 

Finally the scaling properties of the new model are tested by calculating 
the scaling exponent of the number of spins that does change their state with
time. The value of this exponent has been shown to be $3/8$ for the
original Szanjd model.~\cite{3,4} In Fig. 2 a log-log plot of the evolution of
the number of spins in remaining the same state at time $t$ is shown for 
the original Sznajd model and
for the new model proposed here. Plots of Fig. 2 were obtained from simultaneous simulations of both models, using the same random numbers for update each lattice starting from the same initial condition. 
See figure caption for the parameters used in simulations.
It can be seen that the model proposed here share the same type of scaling features as the original Sznajd model, but the value for the scaling exponent seems to be different.
Although, more detailed simulations would be needed in order to verify exactly this last statement. 

\vspace{0.2cm}
The author appreciate the critical review of
the manuscript by Prof. D. Stauffer.

\newpage
\begin{figure}[htbp]
\centerline{\epsfxsize=12cm \epsfbox{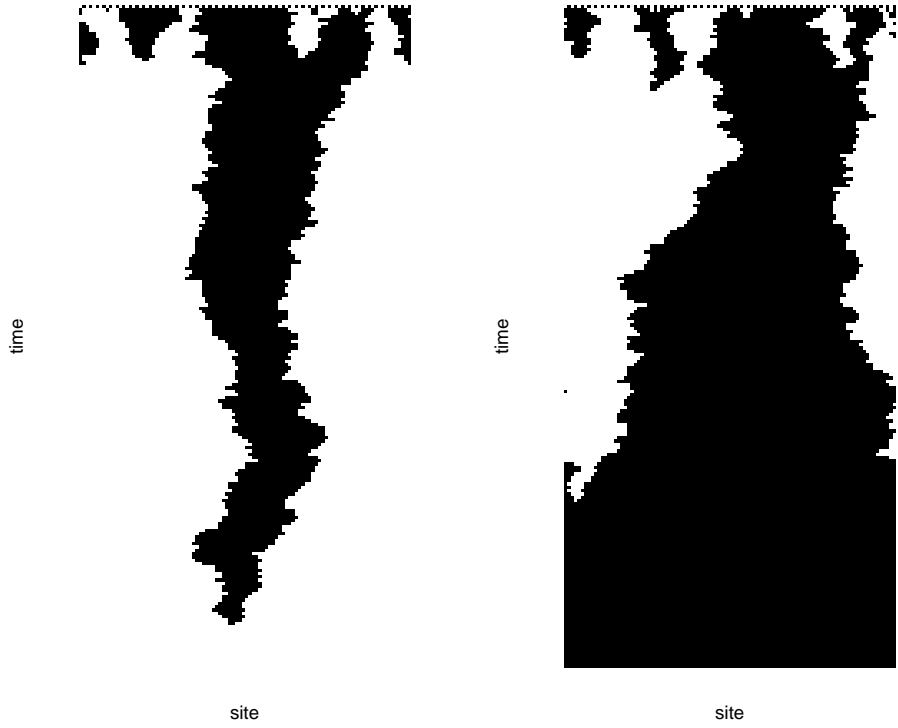}}
\caption{Two sample of time evolution of the modified model using an alternating
state as initial condition.}
\end{figure}

\newpage
\begin{figure}[htbp]
\centerline{\epsfxsize=14cm \epsfbox{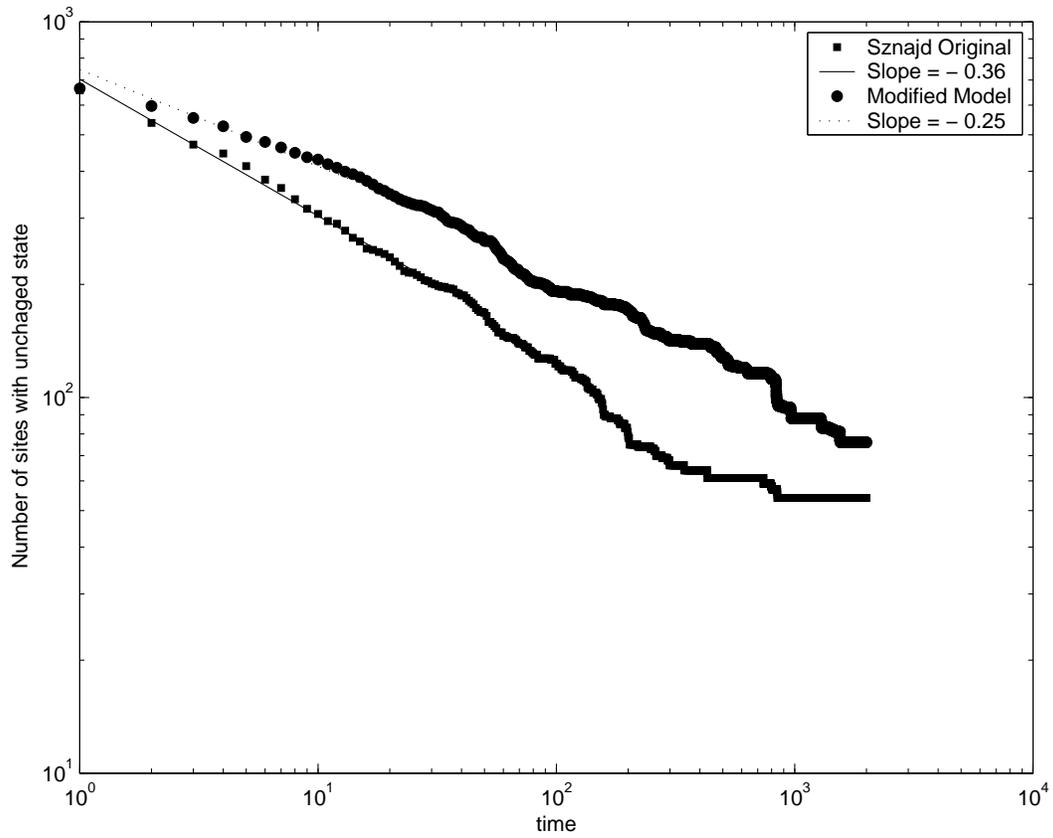}}
\caption{The number of sites with unchanged state follows a power law in both models.
Lattice size N=1000 and total simulation time T=2000.}
\end{figure}

\end{document}